\documentclass{jpsj-suppl}

\title{Photoproduction of $\eta$ and $\eta'$ Mesons on Proton}

\author{V.~L.~Kashevarov$^{1}$, L. Tiator$^{1}$, and M. Ostrick$^{1}$ for A2 Collaboration at MAMI}

\inst{$^{1}$ Institut f\"ur Kernphysik, Johannes Gutenberg-Universit\"at, D-55099 Mainz, Germany}

\email{kashev@kph.uni-mainz.de}

\recdate{September 30, 2016}

\abst{New high-precision total and differential cross sections for $\eta$ and $\eta'$ photoproduction
on the proton obtained by the A2 Collaboration at the Mainz Microtron are presented.
The data for $\eta$ photoproduction demonstrate a cusp at the energy W$\sim$1.9~GeV. 
Furthermore, we present a new version of the $\eta$MAID model for $\eta$ and $\eta'$ photoproduction. 
The model includes 23 nucleon resonances parametrized with Breit-Wigner shapes. 
The background is described by vector and axial-vector meson exchanges in the
$t$ channel using the Regge phenomenology. 
Parameters of the resonances were obtained from a fit to preliminary data of the A2 
Collaboration at MAMI and available data from other collaborations.
The cusp is explained as a threshold effect due to the opening $\eta' p$ decay channel
of the $N(1895)1/2^-$ resonance. }

\kword{meson photoproduction, baryon spectroscopy, partial wave analysis}

\begin{document}
\maketitle

\section{Introduction}
The isobar model $\eta$MAID~\cite{MAID2003} was developed in 2002 for $\eta$ photo-
and electroproduction on nucleons. 
The model includes a nonresonant background, which consists of nucleon Born terms, the 
vector meson exchange in the $t$ channel, and $s$-channel resonance excitations. 
The vector meson contribution is obtained by the $\rho$ and $\omega$ meson exchange in the
$t$ channel with pole-like Feynman propagators. 
The resonance contribution is parametrized by the Breit-Wigner function with energy-dependent width. 
The $\eta$MAID-2003 version describes well the experimental data available in 2002, 
however fails to reproduce the new polarization data obtained in Mainz~\cite{TF_MAMI}.  
An updated version $\eta$MAID-2015~\cite{MAID2015} extended the $\eta$MAID-2003 to higher energies, 
improved a description of the new polarization data, and included the $\eta'$ photoproduction
channel.
In the presented version $\eta$MAID-2016, the background contribution is calculated using Regge
phenomenology. It allowed to describe well the high-energy data. 

\section{Review of experimental data}

The revised model $\eta$MAID-2016 is used for a phenomenological analysis of the data.
Data set includes the latest results for the $\gamma p \to \eta p$ reaction: 
CBELSA/TAPS~\cite{CBELSA_2009} and CLAS~\cite{CLAS_2009} for differential cross sections,
A2MAMI~\cite{TF_MAMI} for T and F, CLAS~\cite{CLAS_2016} for E, GRAAL~\cite{GRAAL_2007} 
for $\Sigma$, and preliminary high statistics data from A2 Collaboration at MAMI for 
the differential cross sections. 
\begin{figure}     
\includegraphics{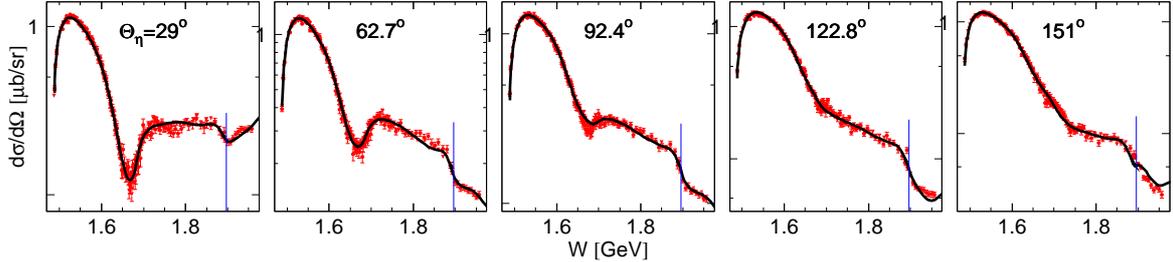}
\caption{Excitation function of $\eta$ photoproduction for selected angular bins. 
The black circles are preliminary A2MAMI data, the red line is the new $\eta$MAID solution, 
the vertical line corresponds to the $\eta'$ threshold. 
}
\label{fig1}
\end{figure}
Energy dependencies of the new differential cross section for selected angular bins
are shown in Fig.~\ref{fig1}. 
The data demonstrate cusp effect, especially at polar angles of $\eta$ meson around 90$^o$.
The differential cross sections cover the energy region from threshold up to W=2.8 GeV.
Polarization observables are from threshold up to W=1.85 GeV for T and F, 2.13 GeV
for E, and 1.91 GeV for $\Sigma$. 
We used also old high energy data~\cite{OLD_Eta} to
determine Regge background contributions. 

Data set for the $\gamma p \to \eta' p$ reaction is much more scarce than for 
$\gamma p \to \eta p$: 
preliminary data from A2 Collaboration at MAMI, CBELSA/TAPS~\cite{CBELSA_2009}, and 
CLAS~\cite{CLAS_2009} for differential cross sections; GRAAL~\cite{GRAAL_2015}
for $\Sigma$.
The differential cross sections cover the energy region from threshold up to W=2.8 GeV.
For beam asymmetry $\Sigma$, data exist only for two energy bins close to threshold.   

\section{$\eta$MAID-2016}

\begin{figure}    
\includegraphics{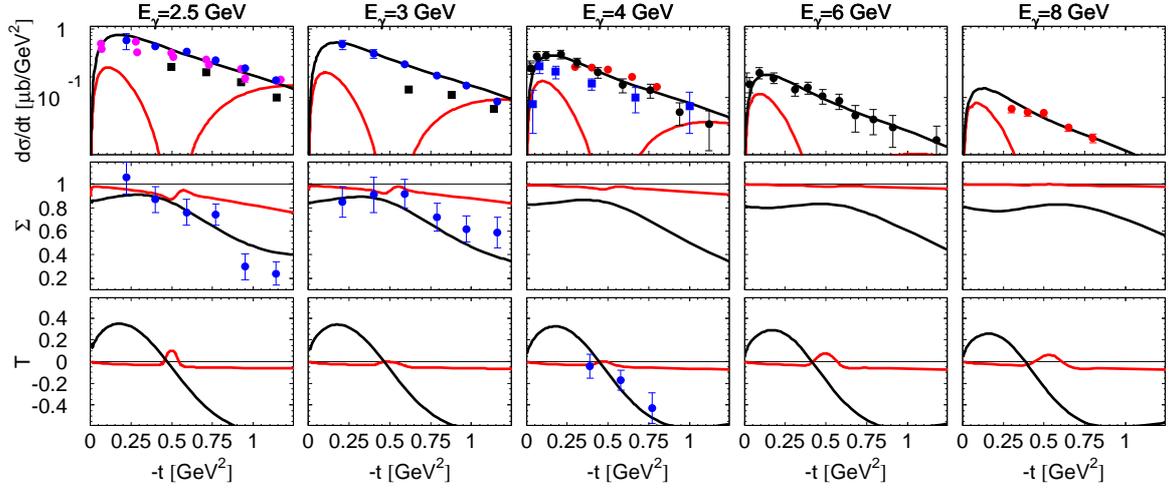}
\caption{Differential cross sections and polarization observables $\Sigma$ and T 
for the $\gamma p \to \eta p$ reaction described by Regge contributions. The red lines are 
the contribution by only $\rho$ and $\omega$ exchange, the black lines show the total 
Regge contribution with the cuts.
The black squares are data from CLAS~\cite{CLAS_2009}, and the other data are given 
in Ref.~\cite{OLD_Eta}.
}
\label{fig2}
\end{figure}

For new $\eta$MAID version, the background from Born terms was excluded because of very small 
contribution.
The pole-like Feynman propagators are replaced by the Regge propagators for each vector meson V in the
t-channel: \\
\begin{equation} \label{eq:rhoregge}
  \frac{1}{t-m_V^2}\ \Longrightarrow\
  \mathcal{P}^V_\mathrm{Regge} =
  \left(\frac{s}{s_0}\right)^{\alpha_V(t)-1}
  \frac{\pi\alpha'_V}{\sin(\pi\alpha_V(t))}
  \frac{\mathcal{S} +e^{-i\pi\alpha_V(t)}}{2}
  \frac{1}{\Gamma(\alpha_V(t))} \,,
\end{equation}
where the parameter $s_0$ is a mass scale taken as $s_0 = 1\;\mathrm{GeV}^2$.
The gamma function $\Gamma(\alpha(t))$ suppresses poles of the propagator
in the unphysical region.
The $\mathcal{S}$ is the signature of the trajectory: $\mathcal{S}$= $(-1)^J$ for bosons, 
so $\mathcal{S}$=-1 for vector mesons. 
In this case, a differential cross section goes to zero if $\alpha(t)$=0 for each trajectory,
see red lines in Fig.~\ref{fig2}.
Using Regge cut phenomenology, Donachie and Kalashnikova~\cite{DoKa} excluded this anomaly for
$\pi^0$ photoproduction. Regge cuts arise from photoproduction rescattering of two Reggions.
Following Ref.~\cite{DoKa}, we used Pomeron with quantum numbers of
the vacuum and tensor meson $f_2$ to produce four cut trajectories: $\rho$P, $\rho f_2$,   
$\omega$P, and $\omega f_2$. 
This is enough to get good description of the differential cross sections at high energies.
Additional exchange by axial-vector meson $b_1$ is needed to describe polarization observables.
All four Regge cuts contribute also to axial-vector exchanges~\cite{DoKa}. 
Unknown coefficients for natural and unnatural parity cuts were obtained by a fit to the data.
Fit result for high energy data for $\gamma p \to \eta p$ reaction is presented in Fig.~\ref{fig2} 
by the black lines. 
The obtained solution describes well the old data~\cite{OLD_Eta}, 
but is not consistent with data from CLAS~\cite{CLAS_2009}.  
Background fixed by this way was extrapolated to the resonance region. 
We used the same fit parameters to determine background contribution for 
the $\gamma p \to \eta' p$ reaction.   

Nucleon resonances in the $s$ channel were parametrized with Breit-Wigner shapes. 
The new model allows the use of 23 resonances
to fit experimental data. The Breit-Wigner mass, total width, branching ratios to $\eta$ and $\eta'$ decays, 
photoexitation helicity couplings $A_{1/2}$ and $A_{3/2}$ are model variable parameters. Branching ratios
to further decay channels, namely $K\Lambda$, $K\Sigma$, $\omega N$, were fixed from PDG~\cite{PDG} or
BnGa analysis~\cite{BnGa12}.  
The new model was fitted to published data of both $\eta$ and $\eta'$ photoproduction on the 
proton~\cite{TF_MAMI, CBELSA_2009, CLAS_2009, CLAS_2016, GRAAL_2007, GRAAL_2015} and to the 
preliminary A2MAMI data.

\begin{figure}    
\includegraphics{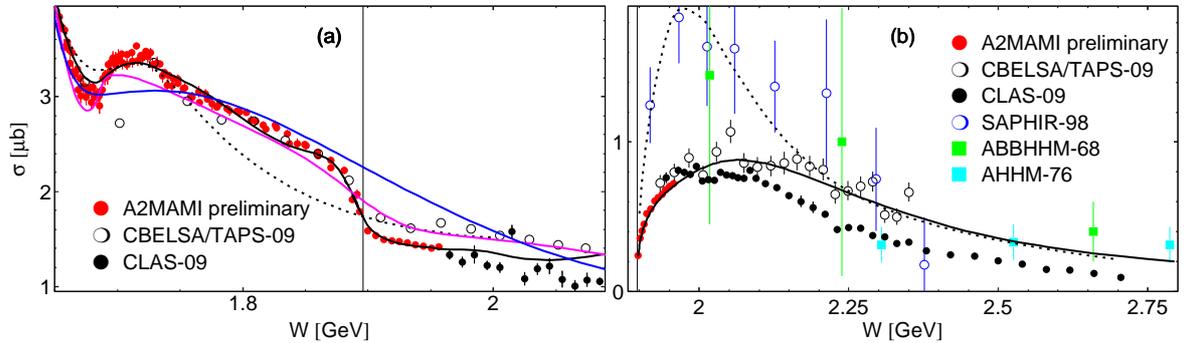}
\caption{Total cross sections for $\gamma p\to \eta p$ (a) and $\gamma p\to \eta' p$ (b).
The new $\eta$MAID solution is shown in the black solid lines, the black dashed lines are 
$\eta$MAID-2003~\cite{MAID2003} (a) and
$\eta$MAIDregge-2003~\cite{MAID2003r} (b) predictions.
Other predictions: SAID-GE09~\cite{SAID} (blue) and
BG2014-2~\cite{BnGa12} (magenta).
 }
\label{fig3}
\end{figure}
\begin{figure}    
\begin{minipage}[t]{0.5\textwidth}
\includegraphics[width=\textwidth]{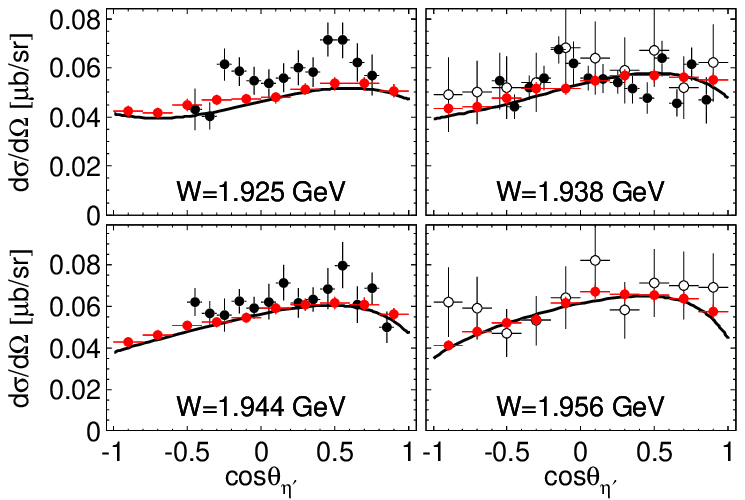}
\caption{New $\eta$MAID solution for the selected \\
energy bins of the $\gamma~p~\to~\eta'~p$ differential cross \\ 
section. Data: preliminary A2MAMI (red), CLAS~\cite{CLAS_2009} (black), 
and CBELSA/TAPS~\cite{CBELSA_2009} (open circles). 
}
\label{fig4}
\end{minipage}
\begin{minipage}[t]{0.5\textwidth}
\begin{center}
\includegraphics[width=0.68\textwidth]{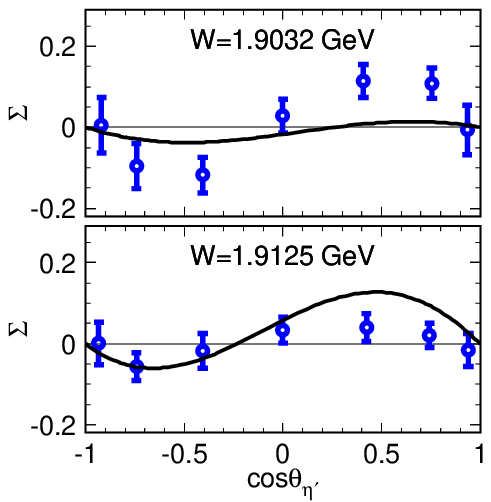}
\caption{New $\eta$MAID solution for the 
$\gamma~p~\to~\eta'~p$ beam asymmetry. 
Data: GRAAL~\cite{GRAAL_2015}.}
\end{center}
\label{fig5}
\end{minipage}
\end{figure}

In Fig.~\ref{fig3}(a), the total $\gamma p \to \eta p$ cross sections 
are shown for the most interesting energy region, 
where differences between the model calculations are especially visible.
The new  $\eta$MAID solution is compared to $\eta$MAID-2003~\cite{MAID2003}, SAID-GE09~\cite{SAID}, and
BG2014-2~\cite{BnGa12} predictions.
The new data demonstrate a strong cusp at an energy corresponding
to $\eta'$ threshold (vertical line in Fig.~\ref{fig5}(a)).
In Fig.~\ref{fig3}(b) the new  $\eta$MAID solution for the total cross section of the $\gamma p \to \eta' p$ 
reaction is compared to $\eta$MAIDregge-2003~\cite{MAID2003r} calculation, which was obtained with
a fit to the old data (SAPHIR-98, ABBHHM-68, AHHM-76)~\cite{OLD_EtaPr}. 
The total cross sections for the CLAS Collaboration, we obtained from differential cross sections 
~\cite{CLAS_2009} using the Legendre fit and are shown for a qualitative comparison.
The total cross sections themselves were not fitted, we show the result of the partial wave analysis.

A key role in the description of the investigated reactions is played by three $s$-wave resonances 
N(1535)$1/2^-$,
N(1650)$1/2^-$, and N(1895)$1/2^-$. The first two give the main contribution to the total cross section
and are known very well. The third of them has only 2-star overall status according to the PDG 
review~\cite{PDG}.
But we have found that namely this resonance is responsible for the cusp effect in the $\eta$ photoproduction 
and provides fast increase of the total cross section in the $\gamma p \to \eta' p$ reaction near the 
threshold.

The results for the $\gamma p \to \eta' p$ reaction are shown for differential cross sections in 
Fig.~\ref{fig4} for selected energy bins and the beam asymmetry $\Sigma$ in Fig.~\ref{fig5}. 
The new $\eta$MAID solution very well describes the new data for the differential cross sections. 
The beam asymmetry $\Sigma$ 
is reproduced in its shape of the angular dependence. However, the energy dependence is inverted.

\section{Summary}
A new reggetized model for $\eta$ and $\eta'$ photoroduction on nucleons
was presented. At energies below W=2.5 GeV nucleon resonance exitation dominate. To describe the data 
in this region we increased the number of $N^*$ resonances from 8 to 23, where
5 of them give only small contributions. At high energies Regge trajectories
of $\rho, \omega, b_1$  and Regge cuts of $\rho$-P, $\omega$-P, $\rho$-$f_2$, $\omega$-$f_2$
were used. The obtained solution describes the data very well up to $E_{\gamma}$=8 GeV.
The cusp in the total cross section of $\gamma$p $\rightarrow$ $\eta$p is explained as
a threshold effect due to the opening of the $\eta' p$ decay channel of the $N(1895)1/2^-$ resonance.


\end{document}